\documentclass[prl,twocolumn,showpacs,amsmath,amsfonts]{revtex4}
\usepackage{bm}
\usepackage{graphicx}

\newcommand{\bra}[1]{\langle #1 | \,}
\newcommand{\ket}[1]{\, | #1 \rangle}

\newcommand{\la}{\lambda}
\newcommand{\om}{\omega}
\newcommand{\Om}{\Omega}
\newcommand{\ga}{\gamma}
\newcommand{\Ga}{\Gamma}
\newcommand{\de}{\delta}
\newcommand{\De}{\Delta}

\begin{document}

\title{Scalable solid-state quantum processor using subradiant two-atom 
states}

\author{David Petrosyan}
\author{Gershon Kurizki}
\affiliation{Department of Chemical Physics, 
Weizmann Institute of Science, Rehovot 76100, Israel}

\date{\today}

\begin{abstract}
We propose a realization of a scalable, high-performance quantum processor 
whose qubits are represented by the ground and subradiant states of effective 
dimers formed by pairs of two-level systems coupled by resonant dipole-dipole 
interaction. The dimers are implanted in low-temperature solid host material 
at controllable nanoscale separations. The two-qubit entanglement either 
relies on the coherent excitation exchange between the dimers or is mediated 
by external laser fields. 
\end{abstract}

\pacs{03.67.Lx, 42.50.Fx}

\maketitle

The main stumbling blocks en route to the realization of useful quantum 
computers, comprised of many qubits, are \cite{RevSteane}: 
(i) {\em fidelity loss} due to decoherence, which grows with the amount 
of single- and two-qubit operations and requires large redundancy for the 
application of error-correction methods; 
(ii) {\em scalability} of the quantum processor (QP), which restricts the 
choice of candidate systems and gives preference to solid-state structures. 
QP proposals and realizations have thus far predominantly involved optical 
manipulations of atoms in ion traps \cite{CZ,MS,SKKLMMRTIWM}, high-Q cavities 
\cite{PGCZ}, and optical lattices \cite{BCJD}. Yet, the decoherence caused by 
radiative (spontaneous emission) and nonradiative processes, as well as 
difficulties with the scalability, cast doubts on the suitability of these 
schemes for truly large-scale quantum computation \cite{BKMPV}. Solid-state 
QP realizations \cite{LDV,Kane,BHHFKWSF,LH} appear to be more promising, 
both principally and technologically.

Here we propose a combined optical/solid-state approach that can significantly 
enhance the speed, fidelity and scalability of a QP. The crux of this approach 
is the hitherto unexplored concept of a ``subradiant dimer'' (SD) qubit: two 
similar two-level systems (atoms or quantum dots) that are separated by a few 
nanometers and interact via the resonant dipole-dipole interaction (RDDI) 
\cite{DDI}, thereby forming an effective ``dimer'', whose ground and 
subradiant (``dark'') states serve as the qubit basis. {\em All the basic 
ingredients of quantum computation} (state preparation, universal logic gates 
and qubit readout) \cite{DiVincenzo} are shown to be realizable by high-speed
optical manipulations of these dimers with very small error probability,
due to strong inhibition of radiative decay. A scalable QP is envisioned in 
a low-temperature solid host material doped with such dimers at controllable 
nanoscale separations.

\begin{figure}[t]
\centerline{\includegraphics[width=8.5cm]{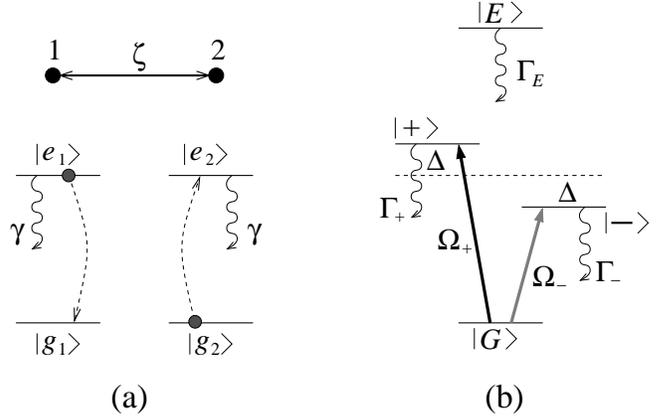}}
\caption{(a) Two TLAs 1 and 2, separated by normalized distance $\zeta$,
interact via RDDI and exchange a single excitation.
(b) Energy level diagram of the resulting ``dimer'' states of the system.
\label{fig:dd_int}}
\end{figure}

Let us recall the cooperative properties of two identical two-level atoms 
(TLAs), 1 and 2, at fixed positions $\textbf{r}_1$ and $\textbf{r}_2$, whose 
ground and excited states are labeled as $\ket{g_{1,2}}$ and $\ket{e_{1,2}}$, 
respectively [Fig. \ref{fig:dd_int}(a)]. The effective (non-Hermitian) 
Hamiltonian of the system can be cast in a form \cite{DDI} 
\begin{equation}
H = H_{\text{A}}+V_{\text{RDDI}} \label{ham},
\end{equation}
where $H_{\text{A}} = \hbar (\om_{eg} - i \ga/2) 
(\ket{e_1} \bra{e_1} + \ket{e_2} \bra{e_2})$ represents the atomic Hamiltonian,
with $\om_{eg}$ being the resonant frequency and $\ga$ the radiative decay rate
on the atomic transition $\ket{e}\to \ket{g}$, and 
$V_{\text{RDDI}} = \hbar \left(\De - i \ga_{12}/2 \right)
(\ket{e_1g_2} \bra{g_1e_2} + \ket{g_1 e_2} \bra{e_1 g_2})$ 
describes the interatomic RDDI potential, whose real part $\De$ is equal to 
the rate of coherent excitation exchange (hopping) between the atoms, and 
the imaginary part $\ga_{12}$ is responsible for the cooperative radiative 
decay of the system. Both $\De$ and $\ga_{12}$ are functions of the 
normalized distance between the atoms $\zeta = q r_{12}$, with 
$q = \om_{eg}/c$ and $r_{12} = |\textbf{r}_1 - \textbf{r}_2|$. 
The diagonalization of Hamiltonian (\ref{ham}) yields the ``dimer'' 
eigenstates $\ket{G} = \ket{g_1 g_2}$, 
$\ket{\pm} = \frac{1}{\sqrt{2}}(\ket{e_1 g_2} \pm \ket{g_1 e_2})$, and
$\ket{E} = \ket{e_1 e_2}$, with the energy eigenvalues 
$\la_G = 0$, $\la_{\pm} = \om_{eg} \pm \De -i \Ga_{\pm}/2$, and
$\la_E = 2 \om_{eg} -i \Ga_E/2$, respectively [Fig. \ref{fig:dd_int}(b)]. 
At small separations $\zeta \ll 1$, the symmetric $\ket{+}$ and doubly-excited 
$\ket{E}$ eigenstates are {\em superradiant}, having the corresponding decay 
rates $\Ga_+ = \ga+\ga_{12} \approx \Ga_E = 2 \ga$, while the antisymmetric 
eigenstate $\ket{-}$ is {\em subradiant}, with the decay rate 
$\Ga_- = \ga - \ga_{12} \approx \ga \zeta^2/5 \ll \ga$ 
\cite{DDI}. The energy levels of states $\ket{\pm}$ are then 
shifted from that of state $\ket{e}$ by $\pm \De$, with 
$|\De| \approx 3 \ga/(4 \zeta^3) \gg \ga$.

The coupling strength of a laser field $\mathcal{E}$, having frequency 
$\om \sim \om_{eg}$ and wave vector $\textbf{k}$, with the dimer is expressed
by its Rabi frequencies, which are equal to $\pm \Om_{-}$ on the transitions 
$\ket{G} \to \ket{-}$ and $\ket{-} \to \ket{E}$, respectively, and to 
$\Om_{+}$ on the transitions $\ket{G} \to \ket{+}$ and $\ket{+} \to \ket{E}$, 
where $\Om_{\pm} = 2^{-1/2} \Om [1 \pm e^{-i \textbf{k} \textbf{r}_{12}}]$ and
$\Om = \mu \mathcal{E}/\hbar$ is the Rabi frequency of the field for a single 
isolated atom, with $\mu$ being the dipole matrix element for the atomic 
transition $\ket{g} \to \ket{e}$. In the limit of small interatomic 
separations, $\Om_+ \simeq 2^{1/2} \Om$ and 
$\Om_- \simeq i 2^{-1/2} \Om \zeta \cos \phi$, where $\phi$ is the angle 
between the vectors $\textbf{k}$ and $\textbf{r}_{12}$. Hence, $\Om_-$ 
identically vanishes if the propagation direction of the field is perpendicular
to the interatomic axis, $\textbf{k} \perp \textbf{r}_{12}$, while it is 
maximized in the $\textbf{k} \parallel \textbf{r}_{12}$ configuration, for 
$\zeta \ll 1$. In physical terms, the subradiant $\ket{G} \to \ket{-}$ 
transition exhibits a {\em quadrupolar} behaviour and dipole-moment 
suppression, due to destructive interference of the two-atom interactions 
with the field, as opposed to their constructive interference in the 
superradiant $\ket{G} \to \ket{+}$ transition. 

Now we are in a position to introduce the concept of the ``subradiant dimer'' 
(SD) qubit. The two qubit states correspond to the ground $\ket{G}$ and 
subradiant $\ket{-}$ states of the dimer. An arbitrary single-qubit operation 
(rotation) can be performed by the laser field $\mathcal{E}_r$ with wave 
vector $\textbf{k}_r \parallel \textbf{r}_{12}$ and frequency 
$\om_r = \om_{eg} -\De$ that is resonant with the qubit transition 
$\ket{G} \to \ket{-}$ [Fig. \ref{fig:dd_int}(b)]. During the qubit flip-time 
$T_{\text{flip}} = \pi / (2 |\Om_-^{(r)}|)$, the probability of error 
$P_-^{\text{sp}}$ due to spontaneous emission from the subradiant state 
$\ket{-}$ has the upper bound 
$P_-^{\text{sp}} \leq \Ga_- T_{\text{flip}}=\pi \ga \zeta/(5 \sqrt{2} \Om_r)$,
while the probability of error due to population transfer from the ground 
state $\ket{G}$ to the superradiant state $\ket{+}$ satisfies 
$P_+^{\text{tr}} \leq \Ga_+ |\Om_+^{(r)}|^2 T_{\text{flip}} /(2 \De)^2 = 
8\sqrt{2} \pi \Om_r \zeta^5/(9 \ga)$. As an example, for the parameters 
$\zeta \simeq 0.02$ and $\Om_r/\ga \simeq 30$, the decay rate of the 
antisymmetric state is $\Ga_- \approx 8 \times 10^{-5} \ga$ and the
error probabilities during the flip-time of a SD qubit are 
$P_-^{\text{sp}} \simeq 3 \times 10^{-4}$ and 
$P_+^{\text{tr}} \ll P_-^{\text{sp}}$, as compared to the corresponding
error probability for a single atom, 
$P_{\text{atom}}^{\text{sp}} \leq \pi \ga/(2 |\Om_r|) \simeq 0.05$. 
Such small errors of the SD qubit are amenable to error correction 
\cite{RevSteane}. 

In order to {\em read-out} (measure) the state of the qubit, we may use a 
modification of the electron-shelving technique \cite{QJumps}. 
Let us apply for a time $T_{\text{rout}}$ a probe field 
$\mathcal{E}_p$ at a frequency $\om_p = \om_{eg} + \De$ that is resonant with 
the dimer transition $\ket{G} \to \ket{+}$. Since the Rabi frequency 
on that transition is much larger than on the qubit transition 
$\ket{G} \to \ket{-}$, from which the probe field is detuned by $2\De$, 
the presence or absence of fluorescence from $\ket{+}$ would indicate 
whether the qubit state is $\ket{G}$ or $\ket{-}$, respectively. However, 
since the frequency $\om_p$ exactly matches that of the transition 
$\ket{-} \to \ket{E}$ [Fig. \ref{fig:dd_int}(b)], the dimer in state 
$\ket{-}$ can first be excited to $\ket{E}$ by absorbing a probe photon, 
then decay to $\ket{+}$, subsequently producing the same fluorescence signal 
as if it were initially in state $\ket{G}$. Therefore, for a reliable 
measurement, the condition $\ga_{-+} T_{\text{rout}} < 1$ should be satisfied, 
where $\ga_{-+} =|\Om_p|^2 \zeta^2/\ga$ is the rate of transition 
$\ket{-} \to \ket{+}$. This leads to the condition 
$\Om_p /\ga < \sqrt{2 \eta} /\zeta$, where $\eta <1$ is the detector 
efficiency. With $\eta \simeq 0.3 $, $\Om_p /\ga \simeq 5$ and 
$\zeta \simeq 0.02$, we obtain 98\% measurement reliability. If, however, 
the probe laser is applied for a time $T_{\text{rout}} \geq \ga_{-+}^{-1}$, 
it will {\em initialize} the state of the qubit to its ground state $\ket{G}$.

\begin{figure}[t]
\centerline{\includegraphics[width=8.5cm]{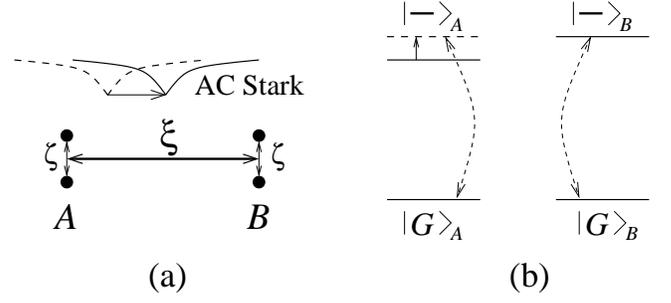}}
\caption{(a) Dimers  $A$ and $B$ are separated by normalized distance
$\xi > \zeta$. An external ac Stark field can switch on and off the RDDI 
between the dimers.
(b) When the qubit transitions of dimers $A$ and $B$ are brought to resonance,
they start swapping a single excitation.
\label{fig:cswp}}
\end{figure}

We next consider the RDDI-induced entanglement between two neighboring 
dimers of size $\zeta$, labeled as $A$ and $B$, whose normalized 
separation $\xi = q r_{AB}$ satisfies the condition $\zeta < \xi \ll 1$ 
[Fig. \ref{fig:cswp}(a)]. The rate of coherent excitation exchange between
the dimers on the qubit transitions $\ket{G}_{A,B} \to \ket{-}_{A,B}$
is given by 
$\De_{AB}^{(-)} \simeq 3 \Ga_- /(4 \xi^3) = 3 \ga \zeta^2/(20 \xi^3)$. 
If the difference in the qubit transition frequencies of the two dimers 
exceeds $\De_{AB}^{(-)}$ (as is usually the case in a solid host), 
then their excitation exchange is effectively switched off. To switch 
their interaction on, one can apply an off-resonant, intense,
standing-wave field, such that dimers $A$ and $B$ are exposed to different 
field amplitudes and therefore undergo different ac Stark shifts 
[Fig. \ref{fig:cswp}(a)]. The standing-wave pattern is then shifted along the 
$A-B$ axis until the qubit 
transitions of the two dimers become resonant. Then, during the time 
$T_{\textsc{swap}} = \pi /(2 \De_{AB}^{(-)})$, the \textsc{swap} 
transformation takes place, 
$\ket{-}_{A(B)}\ket{G}_{B(A)} \to -i \ket{G}_{A(B)}\ket{-}_{B(A)}$, while 
other initial states of the two qubits, $\ket{-}_{A}\ket{-}_{B}$ and 
$\ket{G}_{A}\ket{G}_{B}$, remain unaffected [Fig. \ref{fig:cswp}(b)]. 
In the same way, one can realize the {\em square-root of swap} 
($\sqrt{\textsc{swap}}$) gate between two qubits. By switching on the 
interaction for time $T_{\sqrt{\textsc{swap}}} = \pi /(4 \De_{AB}^{(-)})$, 
one fully entangles the two qubits, attaining an equally-weighted 
superposition of \textsc{swap} and no-\textsc{swap},
\begin{eqnarray}
& & \ket{-}_{A(B)}\ket{G}_{B(A)} \to  \nonumber \\
& & \;\;\; \;\;\;\; \frac{1}{\sqrt{2}} [\ket{-}_{A(B)}\ket{G}_{B(A)} -
i \ket{G}_{A(B)}\ket{-}_{B(A)}] 
\label{sqrt_swap} .
\end{eqnarray}
The main source of error in this scheme is the cooperative spontaneous decay 
of the excited states of the qubits, 
$P_{\text{swap}}^{\text{sp}} \leq 2 \Ga_- T_{\textsc{swap}} = 4 \pi \xi^3/3$. 
With inter-dimer separation $\xi \simeq 0.1 \gg \zeta$, this leads to 
$P_{\text{swap}}^{\text{sp}} \leq 4\times 10^{-3}$, which can be taken care of 
by error correction schemes \cite{RevSteane}. 

\begin{figure}[t]
\centerline{\includegraphics[width=8.5cm]{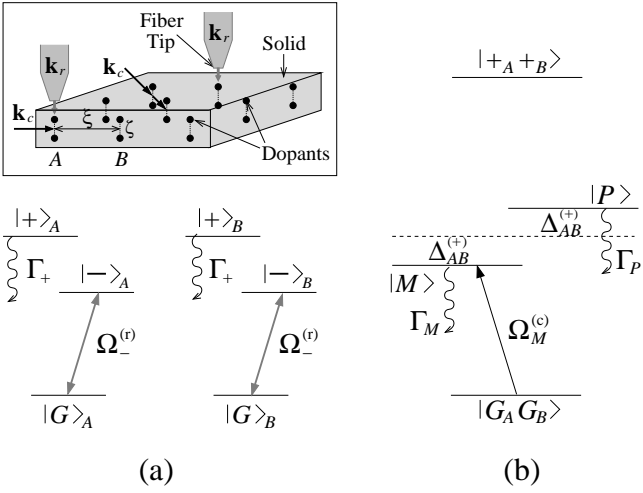}}
\caption{Inset: Schematic drawing of the proposed QP and the geometry for the 
\textsc{cphase} gate: The interatomic axis of each dimer (SD qubit) is 
perpendicular to the inter-dimer axis, 
$\textbf{r}_{12}^{A,B} \perp \textbf{r}_{AB}$. 
Each SD qubit can be separately addressed by a laser field with 
$\textbf{k}_r \parallel \textbf{r}_{12}$. 
Two-qubit interaction is mediated by a coupling field with 
$\textbf{k}_c \parallel \textbf{r}_{AB}$.
(a) Internal level structure of two dimers. 
(b) Eigenstates of the combined system of two dimers.
\label{fig:cphs}}
\end{figure}

A fast {\em controlled-phase} (\textsc{cphase}) logic gate between two closely
spaced SD qubits can be realized by a laser field acting on the auxiliary 
transition $\ket{G} \to \ket{+}$, and thereby populating the state $\ket{+}$. 
This will induce the RDDI between the dimers, causing an excitation 
exchange between state $\ket{+}_A$ of dimer $A$ and state 
$\ket{G}_B$ of dimer $B$ and vice versa [Fig. \ref{fig:cphs}(a)]. From 
the above analysis, the rate of this exchange is given by 
$\De_{AB}^{(+)} \simeq 3 \Ga_+ /(4 \xi^3) = 3 \ga /(2 \xi^3)$, which is much
larger than $\De_{AB}^{(-)}$, since $\Ga_+/\Ga_- \simeq 10/\zeta^2 \gg 1$. 
Therefore, during a time interval that is small compared to 
$|\De_{AB}^{(-)}|^{-1}$, we can neglect the RDDI between the dimers on the 
qubit transitions $\ket{G}_{A,B} \to \ket{-}_{A,B}$ in comparison to that on 
the auxiliary transitions $\ket{G}_{A,B} \to \ket{+}_{A,B}$. To the same
accuracy, the eigenstates of the two-dimer system are $\ket{G_A G_B}$,
$\ket{M} = \frac{1}{\sqrt{2}} (\ket{+_A G_B} - \ket{G_A +_B})$,
$\ket{P} = \frac{1}{\sqrt{2}} (\ket{+_A G_B} + \ket{G_A +_B})$, and 
$\ket{+_A +_B}$. The singly excited states $\ket{M}$ and $\ket{P}$, having 
the decay rates $\Ga_M \simeq \Ga_+ \xi^2/5$ and $\Ga_P \simeq 2 \Ga_+$, 
correspond, respectively, to the antisymmetric and symmetric combinations 
of the superradiant states of the two dimers [Fig. \ref{fig:cphs}(b)]. 
To perform the \textsc{cphase} gate, we irradiate the system with 
the coupling field $\mathcal{E}_c$ having the wave vector 
$\textbf{k}_c \parallel \textbf{r}_{AB}$ and frequency 
$\om_c = \om_{eg} + \De - \De_{AB}^{(+)}$ that is resonant with the transition 
$\ket{G_A G_B} \to \ket{M}$ (Fig. \ref{fig:cphs}-inset). The Rabi frequencies 
of this field on the transitions $\ket{G_A G_B} \to \ket{M}$ and 
$\ket{G_A G_B} \to \ket{P}$ are equal, respectively, to 
$\Om_M^{(c)} = \Om_c \xi$ and $\Om_P^{(c)} = 2 \Om_c$.
Since $\textbf{k}_c \perp \textbf{r}_{12}^{A,B}$, this field does not
couple to the qubit transitions of the dimers. During the time 
$T_{\textsc{cphase}} = \pi/\Om_M^{(c)}$, the system of two dimers, being 
initially in the state $\ket{G_A G_B}$, undergoes the Rabi cycle from
$\ket{G_A G_B}$ to $\ket{M}$ and back, resulting in the $\pi$-phase-shift
\begin{equation}
\ket{G_A G_B} \to - \ket{G_A G_B}. \label{cphase}
\end{equation} 
This transformation corresponds to the \textsc{cphase} logic gate, 
since all other initial states, such as $\ket{-}_{A}\ket{-}_{B}$ and 
$\ket{-}_{A(B)}\ket{G}_{B(A)}$, remain unaffected, due to the 
fact that the RDDI between the dimers is present only if their combined 
state is either $\ket{G}_{A}\ket{+}_{B}$ or $\ket{+}_{A}\ket{G}_{B}$,
otherwise the coupling field is off-resonant with the system. 
The error during this gate operation is due to the spontaneous emission from 
the state $\ket{M}$, with the probability
$P_{\textsc{cphase}}^{\text{sp}} \leq \Ga_M T_{\textsc{cphase}}=
2 \pi \ga \xi/5 \Om_c$, as well as due to population transfer from the 
state $\ket{G_A G_B}$ to the state $\ket{P}$, with the probability 
$P_{\textsc{cphase}}^{\text{tr}} \leq \Ga_P |\Om_P^{(c)}|^2 
T_{\textsc{cphase}} /(2 \De_{AB}^{(+)})^2 = 
16 \pi \Om_c \xi^5/(9 \ga)$. With $\Om_c/\ga \simeq 30$ 
and $\xi \simeq 0.1 > \zeta$, we obtain 
$P_{\textsc{cphase}}^{\text{sp}} \leq 4 \times 10^{-3}$ and 
$P_{\textsc{cphase}}^{\text{tr}} \ll P_{\textsc{cphase}}^{\text{sp}}$.
This error probability is exactly the same as for a \textsc{swap} gate
with similar $\xi$, but the \textsc{cphase} gate is then 25 times faster 
than the \textsc{swap} gate, since 
$T_{\textsc{swap}}/T_{\textsc{cphase}} = 10 \Om_c \xi^4/(3 \ga \zeta^2)$.

Having established all the basic principles of the proposed QP, we now 
describe its possible realization. We envision a solid-state host 
doped with active atoms having {\em non-degenerate} ground state, so as 
to avoid mixing of various degenerate atomic states, which would invalidate 
our simple two-level atomic model. Possible candidate systems include 
sulphur-doped silicon, rare-earth (Yb or Nd) doped crystals \cite{rearth}, 
or semiconductor based nanostructures (quantum dots) \cite{QDots}. The 
implantation of dopants and dots with controllable separations of few 
nanometers is achievable with reasonable accuracy \cite{nano,BHHFKWSF}.

With the arrangement of dopants shown in Fig. \ref{fig:cphs}-inset, 
our scheme is capable of implementing arbitrary one-qubit rotations and 
two-qubit logic gates, so as to obtain any desired unitary transformation 
\cite{RevSteane}. 
{\em (a) Individual SD qubits} would be rotated or read-out (and initialized) 
by laser fields with frequency $\om_r$ or $\om_p$, respectively (see above), 
and wave vector parallel to the interatomic axis, using the ``near-field'' 
technique (Fig. \ref{fig:cphs}-inset). The polarization of these fields can be 
chosen such that they act only on the atomic transition from the nondegenerate
ground state to one of the magnetic sublevels of the excited state, 
consistently with our two-level description of the atoms. 
{\em (b) The \textsc{cphase} gate} between a chosen pair of qubits $A$ and $B$ 
is executed by a coupling field with frequency $\om_c$ and wave vector 
$\textbf{k}_c \parallel \textbf{r}_{AB}$ that are specific for that pair. 
{\em (c) The \textsc{swap} action} between neighboring qubits can be used to 
convey the information in the QP, step-by-step, over large distances for which 
the direct RDDI vanishes. To neutralize the \textsc{swap}, one can flip the 
qubits at time intervals short compared to $[\De_{AB}^{(-)}]^{-1}$, which 
is equivalent to the spin echo technique used in NMR \cite{NMR}. 
Alternatively, the $\sqrt{\textsc{swap}}$ gate between two qubits $A$ and $B$ 
can be switched on and off via external ac Stark fields. 

Throughout this paper we have only dealt with the {\em radiative relaxation} 
of the excited atomic state $\ket{e}$. This is adequate provided the competing 
{\em nonradiative relaxation processes} are strongly {\em suppressed} by 
working below the liquid helium temperature \cite{phonons} and/or using fast 
ac Stark modulation of the vibrationally relaxing levels \cite{zeno}. 
Another important consideration is the {\em inhomogeneous broadening} of the 
atomic resonances. Consider two atoms having slightly different resonant 
frequencies, $\om_{eg}^{(2)} - \om_{eg}^{(1)} = \de$, due to the host 
inhomogeneity. This frequency mismatch results in an increase of the decay 
rate $\Ga_-$ of the SD qubit in the amount $\ga \de^2/(8 \De^2)$. If we 
require that this additional relaxation rate does not exceed $\Ga_-$ for 
two resonant atoms, we obtain that the inhomogeneous width $\de$ must be
less than $\ga/\zeta^2$, which, for $\zeta \simeq 0.02$, yields
$\de\leq 2.5 \times 10^{3} \ga$.

It is instructive to compare our scheme with previously considered 
optically-controlled single- and two-qubit quantum gates:

1) In a commonly used optical scheme \cite{BCJD,PGCZ,LH}, 
a Raman qubit is represented by two metastable ground states $\ket{g_1}$ 
and $\ket{g_2}$ that are manipulated by two laser fields detuned by the amount 
$\de_e \gg \ga_e$ from the intermediate excited state $\ket{e}$ having the 
spontaneous decay rate $\ga_e$. The error probability during the qubit flip 
$P_e^{\text{sp}} \leq \pi \ga_e /(2\de_e)$ is then an order of magnitude 
larger than for the SD qubit, given {\em similar} values of the single-photon 
$\Om_R$ and the effective two-photon $\Om_R^2/\de_e$ Rabi frequencies.

2) A \textsc{cphase} logic gate between two closely spaced Raman qubits
[see 1) above], $A$ and $B$, trapped in an optical lattice \cite{BCJD}, is 
realized by an off-resonant ``catalysis'' field with Rabi frequency $\Om_C$,
which induces a RDDI-dependent ac Stark shift of the two-qubit states. One 
then can show that during the gate operation, the probability of error due to 
spontaneous decay of the excited states $\ket{e}_{A,B}$ is given by 
$P_{\textsc{cphase}}^{(R)} \simeq 8 \pi \xi^3/3$, 
where $\xi$ is the Lamb-Dicke parameter. With $\xi \simeq 0.1$, we obtain that 
$P_{\textsc{cphase}}^{(R)} \simeq 8 \times 10^{-3}$, which is 
twice worse than in our scheme with the same $\xi$. More dramatically, for 
{\em similar} field strengths, e.g., $\Om_C/ \ga_e \simeq 30$ and 
$\de_e \simeq 5 \De_{AB}^{(R)} \gg \Om_C$, where $\De_{AB}^{(R)}$ is 
the RDDI coupling strength between the atoms on the transitions 
$\ket{g_2}_{A,B} \to \ket{e}_{A,B}$, we find that the SD qubit 
implementation of the \textsc{cphase} gate is $\sim 30$ times faster.

3) A \textsc{cphase} gate in an ion trap \cite{CZ} operates with speed
and error probability similar to our scheme. The error in the ion trap QP 
is caused by the radiative decay of the auxiliary excited state, but one must 
also reckon with error due to the phonon-mode decoherence 
\cite{MS,SKKLMMRTIWM}. The main limitations of ion trap schemes are related 
to difficulties with their scalability.

To conclude, our proposal for an optically-manipulated,  solid-state quantum 
processor has no principal limitations on scalability. It allows us to 
suppress radiative decoherence and enhance the speed of photon-mediated 
quantum-logic gates, owing to the use of the ground and subradiant states 
of effective dimers formed by resonant dipole-dipole interacting two-level 
systems. These states constitute a physically realistic, simple and robust 
``decoherence-free subspace'' \cite{DFS}, whose implementation draws 
efficiently upon the system resources (only two atoms per qubit). The highly 
challenging experimental realization of such a quantum computer requires 
nanofabrication techniques with nanometer precision of dopant or quantum 
dot implantation \cite{nano,BHHFKWSF}.



\end{document}